\documentclass[aps,prl,twocolumn,superscriptaddress,showpacs,draft]{revtex4}
\input{epsf}

\newcommand{\rem}[1]{}

\begin{document}

\title{Dissipative quantum chaos: transition from wave packet 
collapse to explosion}
\author{Gabriel G. Carlo}
\affiliation{Center for Nonlinear and Complex Systems, Universit\`a degli 
Studi dell'Insubria and Istituto Nazionale per la Fisica della Materia, 
Unit\`a di Como, Via Valleggio 11, 22100 Como, Italy}
\author{Giuliano Benenti} 
\affiliation{Center for Nonlinear and Complex Systems, Universit\`a degli 
Studi dell'Insubria and Istituto Nazionale per la Fisica della Materia, 
Unit\`a di Como, Via Valleggio 11, 22100 Como, Italy}
\affiliation{Istituto Nazionale di Fisica Nucleare,
Sezione di Milano, Via Celoria 16, 20133 Milano, Italy}
\author{Dima L. Shepelyansky}
\affiliation{Laboratoire de Physique Th\'eorique, 
UMR 5152 du CNRS, Universit\'e Paul Sabatier, 
31062 Touluse Cedex 4, France}
\date{Dated: March 8, 2005}


\begin{abstract} 
Using the quantum trajectories approach
we study the quantum dynamics of a dissipative chaotic system
described by the Zaslavsky map. For strong dissipation the 
quantum wave function in the phase space collapses 
onto a compact packet which follows classical chaotic dynamics
and whose area is proportional to the Planck constant.
At weak dissipation
the exponential instability of quantum  dynamics 
on the Ehrenfest time scale dominates and leads to wave
packet explosion. The transition from collapse to explosion
takes place when the dissipation time scale exceeds 
the Ehrenfest time. For integrable nonlinear dynamics
the explosion practically disappears leaving place to collapse. 
\end{abstract}
\maketitle

Technological progress leads to the investigation of physical 
phenomena at smaller and smaller scales, where both quantum and
dissipative effects play a very important role. At present,
general theoretical concepts for the description of
 quantum dissipative systems
are well developed and established \cite{weissbook,ingold}.
A major tool is the master equation, that governs the 
evolution of the density matrix \cite{lindblad}. 
For the simplest dynamics this equation can be solved exactly. 
However, for complex nonlinear systems analytical solution is absent
and even numerical simulations become
very difficult. Indeed, for a system whose Hilbert space has
dimension $N$, one has to store and evolve a density matrix of 
size $N\times N$. In spite of these limitations, numerical simulations
of the master equation allowed to perform the first studies of the quantum 
dynamics of classically chaotic dissipative systems 
showing a quantum strange attractor \cite{graham}.

Quantum trajectories are a very convenient tool to simulate 
dissipative systems \cite{schack,brun}. Instead of direct solution of
the master equation, quantum trajectories allow us to store only a 
stochastically evolving state vector of size $N$.
By averaging over many runs we get the same probabilities 
(within statistical errors) as the ones obtained
through the density matrix directly.
Besides their practical convenience, quantum trajectories also 
provide a good illustration of individual experimental runs 
\cite{dalibard}. Indeed, modern experiments often enable us
to address a single quantum system evolving under the unavoidable 
influence of the environment.

It is known that for linear systems dissipation 
leads to wave packet localization \cite{linear}. 
Numerical results as well as theoretical arguments indicated
that localization can occur also in nonlinear systems
\cite{schack2,percival}.
On the other side, in absence of dissipation
it is known that the instability 
of classical dynamics leads to exponentially fast spreading 
of the quantum wave packet on the logarithmically short
Ehrenfest time scale $t_E \sim |\ln \hbar|/\lambda$ 
\cite{berman,chirikov81}. 
Here $\lambda$ denotes the Lyapunov exponent which gives the rate 
of exponential instability of classical chaotic motion, and 
$\hbar$ is the dimensionless effective Planck constant of the 
system. In this paper we show that for the dissipative quantum chaos
there exist two regimes: one with
the wave packet explosion 
(delocalization) induced by chaotic dynamics and 
another with the wave packet 
collapse (localization) caused by dissipation. 
We argue that the transition (or crossover)
from  collapse to explosion
takes place when the dissipation time $1/\gamma$ 
becomes larger than the Ehrenfest time $t_E$ 
($\gamma$ is the dissipation rate).
		
We investigate the quantum evolution of a kicked system subjected 
to a dissipative friction force. Assuming the Markov approximation, 
we can write a master equation in the Lindblad form
\cite{lindblad}:
\begin{equation}
\dot{\hat{\rho}} = -i
[\hat{H},\hat{\rho}] - \frac{1}{2} \sum_{\mu}
\{\hat{L}_{\mu}^{\dag} \hat{L}_{\mu},\hat{\rho}\}+
\sum_{\mu} \hat{L}_{\mu} \hat{\rho} \hat{L}_{\mu}^{\dag},
\label{lindblad}
\end{equation}
where $\hat{\rho}$ is the density operator,
\{\,,\,\} denotes the anticommutator, 
$\hat{L}_{\mu}$ are the Lindblad operators, 
which model the effects of the environment, and 
$\hat{H}$ is the Hamiltonian of the system.
We consider a kicked system, described by the Hamiltonian
\begin{equation}
\hat{H}=\frac{\hat{n}^2}{2}+
k\cos\,(\hat{x})\sum_{m=-\infty}^{+\infty}\delta(\tau-mT),
\end{equation}
where $T$ is the kicking period and the operators $\hat{x}$
and $\hat{n}=-i(d/dx)$ come from quantizing the classical 
variables $x\in [0,2\pi[$ and $n\in(-\infty,+\infty)$.
This Hamiltonian corresponds to the kicked rotator
\cite{izrailev}, a paradigmatic model in the fields of nonlinear 
dynamics and quantum chaos. This model is also on the focus 
of experimental investigations with cold atoms 
in optical lattices \cite{raizen}.
We assume that dissipation
is described by the lowering operators
\begin{equation}
\begin{array}{l}
\hat{L}_1 = g \sum_n \sqrt{n+1} \; |n \rangle \, \langle n+1|,\\
\hat{L}_2 = g \sum_n \sqrt{n+1} \; |-n \rangle \, \langle -n-1|,
\end{array}
\end{equation}
with $n=0,1,...$ eigenvalues of the operator $\hat{n}$.
At the classical limit, the evolution of the system in one 
period is described by the Zaslavsky map \cite{zaslavsky}
\begin{equation}
\left\{
\begin{array}{l}
n_{t+1}=(1-\gamma)n_t+k\sin x_t,
\\
x_{t+1}=x_t+Tn_{t+1},
\end{array}
\right.
\label{dissmap} 
\end{equation}
where the discrete time $t$ is measured in number of kicks and 
$1-\gamma=\exp(-g^2)$. This map describes a friction force
proportional to velocity. We have $0\le \gamma\le 1$; the limiting 
cases $\gamma=0$ and $\gamma=1$ correspond to Hamiltonian evolution
and overdamped case, respectively. 
Introducing the rescaled momentum variable $p=Tn$,
we can see that the classical dynamics depends only on the
parameters $K=kT$ and $\gamma$.
Since $[\hat{x},\hat{p}]=[\hat{x},T\hat{n}]=iT$, the 
effective Planck constant is $\hbar=T$. 
The classical limit corresponds to $\hbar\to 0$, 
while keeping $K=\hbar k = const$.

The first two terms of Eq.~(\ref{lindblad}) can be regarded as the
evolution governed by an effective non-Hermitian Hamiltonian,
$\hat{H}_{\rm eff}=\hat{H}_s+i\hat{W}$, with
$\hat{W}=-1/2
\sum_{\mu}\hat{L}_{\mu}^{\dag}\hat{L}_{\mu}$. In turn,
the last term is responsible for the so-called quantum jumps.
Taking an initial state $|\phi(\tau_0)\rangle$,
the jump probabilities $dp_{\mu}$ in an infinitesimal time
$d\tau$ are defined by
$
dp_{\mu}\!=\!\langle \phi(\tau_0)| \hat{L}_{\mu}^{\dag}
\hat{L}_{\mu} |\phi(\tau_0)\rangle d\tau,
$
and the new states after the jumps by
$|\phi_{\mu}\rangle = \hat{L}_{\mu}
|\phi(\tau_0)\rangle/||{L}_{\mu} |\phi(\tau_0)\rangle||$.
With probability
$dp_{\mu}$ a jump occurs and the system is left in the state
$|\phi_{\mu}\rangle$. With probability $1-\sum_{\mu} dp_{\mu}$ there
are no jumps and the system evolves according to the effective
Hamiltonian $\hat{H}_{\rm eff}$. In this case we end up with the state
$|\phi_0\rangle =
{(\openone-i H_{\rm eff} dt/\hbar) |\phi(t_0)\rangle}/
{\sqrt{1-\sum_{k} dp_{k}}}$.
We note that the normalization is included
also in this case because the evolution governed by 
$\hat{H}_{\rm eff}$ is non-Hermitian.
To simulate numerically the above described jump picture we 
use the so-called Monte Carlo wave function approach \cite{dalibard}.
The changing state of a single open quantum system is represented
directly by a stochastically evolving quantum wave function, as for
a single run of a laboratory experiment. We say that a single
evolution is a quantum trajectory.

\begin{figure}
\centerline{\epsfxsize=8.5cm\epsffile{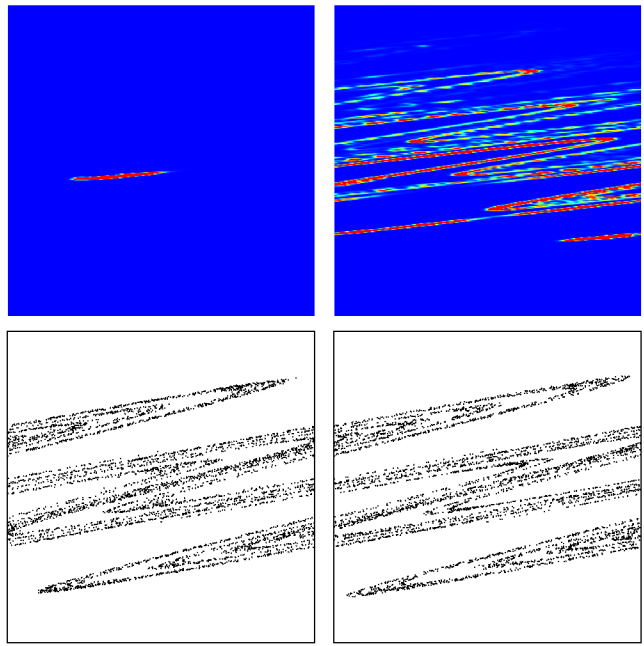}}
\caption{(color online) Top: Husimi functions 
in phase space for a single quantum trajectory 
taken after $t=300$ kicks, at $K=7$, $\hbar=0.012$, $\gamma=0.5$ (left)
and $\gamma=0.01$ (right).
Here the coordinates $x$ (horizontal axis) and $p$ (vertical axis)
vary in the intervals:
$0\le x <2\pi$, $-25\le p \le 25$ (left) and 
$-100\le p \le 50$ (right); the width of the 
$p$-interval is the same in both cases for comparison purposes.
The initial Gaussian wave packet is 
located at $(\langle x \rangle , \langle p \rangle)=(5 \pi /4,0)$. 
The color is proportional to  density: blue for zero and 
red for maximum. 
Bottom: quantum Poincar\'e section (left), obtained from
average quantum $x,p$ values for the case  
of top left panel
and its classical counterpart (right). In these panels $0\le x <2\pi$
and $-15\le p \le 15$.}
\label{figure1}
\end{figure}

We focus first on the chaotic regime for the kicked rotator dynamics.
Therefore, we consider $K=7$, corresponding to a positive Lyapunov 
exponent $\lambda\approx \ln (K/2) =1.25$. 
The localization-delocalization transition is clearly illustrated
in the two top panels of Fig.~\ref{figure1}. They show the Husimi function 
\cite{Husimi} corresponding to a single quantum trajectory evolution,
computed after $t=300$ kicks. In both cases the initial wave packet 
is a Gaussian state with equal uncertainties 
$\Delta x=\Delta p=\sqrt{\hbar/2}$. 
We can see that for strong dissipation ($\gamma=0.5$) the wave 
function of a single quantum trajectory
at $t=300$ is localized in the phase space (top left
panel in Fig.~\ref{figure1}). In contrast, the case of weak 
dissipation ($\gamma=0.01$, top right) is characterized by 
wave packet delocalization.
Since for strong dissipation the wave packet is localized in
phase space, it makes sense to draw a quantum Poincar\'e 
section by printing the expectation values $\langle x \rangle$ and 
$\langle p \rangle$ at each map step. The quantum Poincar\'e 
section is shown in Fig.~\ref{figure1} (bottom left) and is characterized
by the appearance of a strange attractor. A very similar strange attractor
is obtained also from the classical Poincar\'e section corresponding
to the Zaslavsky map (\ref{dissmap}) (see Fig.~\ref{figure1} bottom right).
We also note that the Husimi function obtained in the case of weak 
dissipation exhibits a spreading of the quantum wave packet over the 
strange attractor. Also in the strongly dissipative regime 
the localized wave packet is stretched along the direction of the 
attractor.

\begin{figure}
\centerline{\epsfxsize=8cm\epsffile{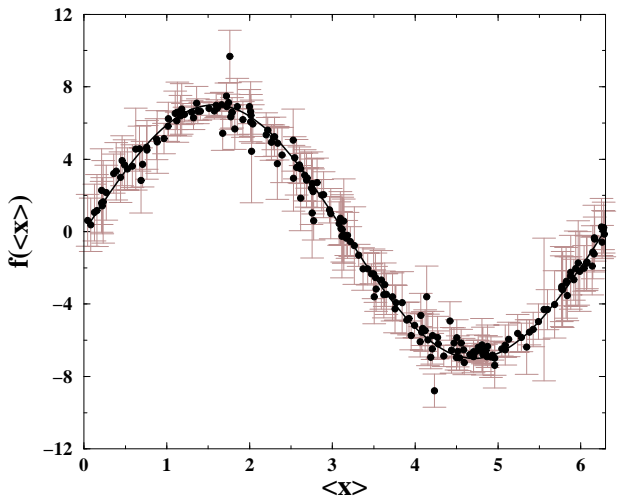}}
\caption{Comparison between $f(\langle x \rangle)$ computed from
the quantum dynamics (circles) and the kick function 
$f(x)=K\sin x$ (solid curve).  
Parameter values and initial conditions are as in Fig.~\ref{figure1} 
(left top and bottom panels). 
Error bars represent the quantum uncertainty in $f(\langle x \rangle)$.}
\label{figure2}
\end{figure}

A further confirmation of the good agreement between the 
classical and quantum dissipative evolutions is obtained by
computing the function 
$f\equiv \langle p \rangle_{t+1}-(1-\gamma) \langle p \rangle_t$
\cite{noted}.
From the classical map (\ref{dissmap}) we expect $f(x)=K\sin x$.
The comparison between $f(\langle x \rangle)$ and 
the function $f(x)$, shown in Fig.~\ref{figure2}, 
indicates that the Zaslavsky map provides a good description of 
the quantum wave packet motion. Indeed the points $f(\langle x \rangle)$
are concentrated around the curve $f(x)$, with dispersion proportional 
to $\sqrt{\hbar}$ (see Fig.~\ref{figure4} below).
Therefore a quantum trajectory exhibits the same important features  
of a classical trajectory, including the exponential instability, with
rate given by the classical Lyapunov exponent. A significant 
difference between quantum and classical trajectories is the presence 
of quantum noise \cite{noise}. Therefore, a more precise identification
can be done between quantum evolution and noisy classical evolution, 
the noise amplitude being $\Delta x \sim \Delta p \propto \sqrt{\hbar}$.  
It is interesting to note that the chaotic behavior of classical systems 
can be reproduced also by non-dissipative continuously
measured quantum systems \cite{habib,milburn,deutsch}.

\begin{figure}
\centerline{\epsfxsize=8.5cm\epsffile{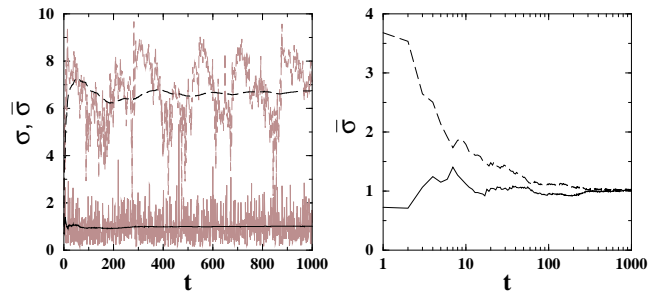}}
\caption{Left: Dispersion $\sigma$ (fluctuating curves) 
and cumulative average $\overline{\sigma}$ (smooth curves) 
as a function of the number of kicks $t$, for the cases 
considered in Fig.~\ref{figure1}, with $\gamma=0.01$
(dashed curves) and $\gamma=0.5$ (full curves).
Right: $\overline{\sigma}$ versus time $t$ at $\gamma=0.5$, 
starting from an initial Gaussian wave packet (full curve,
also shown in the left panel) and a coordinate eigenstate 
$|x\rangle=|x_0\rangle$, with $x_0=\pi$ (dashed curve).} 
\label{figure3}
\end{figure}

The wave packet dispersion is measured by 
$\sigma_t=\sqrt{(\Delta x)_t^2+(\Delta p)_t^2}$. 
This quantity is evaluated, for weak and strong dissipation,
in Fig.~\ref{figure3} (left panel), using the same parameter values 
and initial conditions as in Fig.~\ref{figure1}. In both cases there are
strong fluctuations, which we smooth down by computing the 
cumulative average $\overline{\sigma}_t\equiv \frac{1}{t}\sum_{j=1}^t\sigma_j$.
The convergence of the time averaged quantity $\overline{\sigma}_t$
to a limiting value is clear. It is also evident that the wave packet
spreading is much stronger at weak than at strong dissipation.
We would like to stress that the same limiting value of the 
average dispersion $\overline{\sigma}$ is obtained for any quantum 
trajectory, independently of the initial condition. 
This is demonstrated in the right panel of 
Fig.~\ref{figure3}, where we compare $\overline{\sigma}_t$ 
for two completely different initial conditions: a Gaussian wave
packet and an eigenstate of the operator $\hat{x}$, that is, 
$|x\rangle = |x_0\rangle$. In the latter case there is a complete 
delocalization along $p$ (limited only by the size of the basis 
considered in our numerical simulations) and dissipation leads 
to the collapse of the wave packet, which eventually becomes localized
in phase space.

The average dispersion $\overline{\sigma}$ 
of the wave packet as a function
on the dissipation strength $\gamma$ is shown in Fig.~\ref{figure4}, 
for a few values of the effective Planck constant $\hbar$, 
with $0.012\le \hbar \le 0.33$. The localization-delocalization
transition can be seen for all $\hbar$ values. In Fig.~\ref{figure4}
inset we consider the scaled dispersion 
$\overline{\sigma}_s \equiv \overline{\sigma}/\sqrt{\hbar}$.
At strong dissipation all curves collapse, while at weak 
dissipation the scaling $\overline{\sigma}\propto\sqrt{\hbar}$
is not fulfilled. Our numerical results can be explained as follows.
Due to the exponential instability of chaotic dynamics
the wave packet spreads exponentially and for times 
shorter than the Ehrenfest time we have
$\sigma_t\sim \sqrt{\hbar}\exp(\lambda t)$. 
The dissipation localizes the wave packet on a time scale
of the order of $1/\gamma$. Therefore, for $1/\gamma \ll t_E$, we 
obtain $\overline{\sigma}\sim \sqrt{\hbar} \exp(\lambda/\gamma) \ll 1$. 
In contrast, for $1/\gamma >t_E$ the chaotic wave packet explosion 
dominates over dissipation and we have complete delocalization 
over the angle variable. In addition, in this case,  
the wave packet spreads algebraically due to diffusion for $t > t_E$. 
For $t\gg t_E$ we have
$\sigma_t\sim \sqrt {D(K) t}$, $D(K) \approx  K^2/2$ being 
the diffusion coefficient. This regime continues up to the 
dissipation time $1/\gamma$, so that 
$\overline{\sigma} \sim \sqrt{D(K)/\gamma}$.
According to the above argument, the 
transition from collapse to explosion, which we wish to 
call \emph{Ehrenfest explosion}, takes place at
\begin{equation}
t_E \sim |\ln \hbar|/\lambda \sim 1/\gamma \; .
\end{equation}
Our numerical data at moderate values of $\hbar > 0.01$
indicate a smooth transition (crossover). However,
we cannot exclude from our data that in the limit 
$\hbar \rightarrow 0$ the transition becomes sharp.
Since the dependence on $\hbar$ is only logarithmic
it is difficult to check numerically the above relation.
However, it is compatible with our data obtained for
$\hbar > 0.01$. 
First of all, in the localized regime $\gamma t_E > 1$
the scaling law $\overline{\sigma}\propto \sqrt{\hbar}$ is 
satisfied. Moreover, it is satisfied down to 
smaller and smaller $\gamma$ values when $\hbar$ is reduced. 
Therefore, even for infinitesimal dissipation strengths the 
quantum wave packet is eventually localized when $\hbar\to 0$:
we have $\lim_{\hbar\to 0} \overline{\sigma}=0$. 
In contrast, in the Hamiltonian case ($\gamma=0$)   
$\lim_{\hbar\to 0} \overline{\sigma}=\infty$. 
This result underlines the importance of a (dissipative) 
environment in driving the quantum to classical transition: 
only for open quantum systems the classical concept of trajectory
is meaningful for arbitrarily long times. On the contrary, for
Hamiltonian systems a description based on wave packet trajectories 
is possible only up to the Ehrenfest time scale. 

We would like to emphasize the role played by chaotic 
motion. For this purpose, in Fig.~\ref{figure4} we also show 
$\overline{\sigma}$ as a function of $\gamma$ in the integrable 
regime at $K=0.7$, for $\hbar=0.012$. In this case the wave packet
dispersion is much smaller than in the chaotic regime: the Ehrenfest
time scale is algebraic and not logarithmic in $\hbar$. 
Thus, a much weaker dissipation is sufficient to 
localize the wave function in the case of integrable dynamics. 
This can be clearly seen from our numerical
data shown in Fig.~\ref{figure4}. 

\begin{figure}
\centerline{\epsfxsize=8cm\epsffile{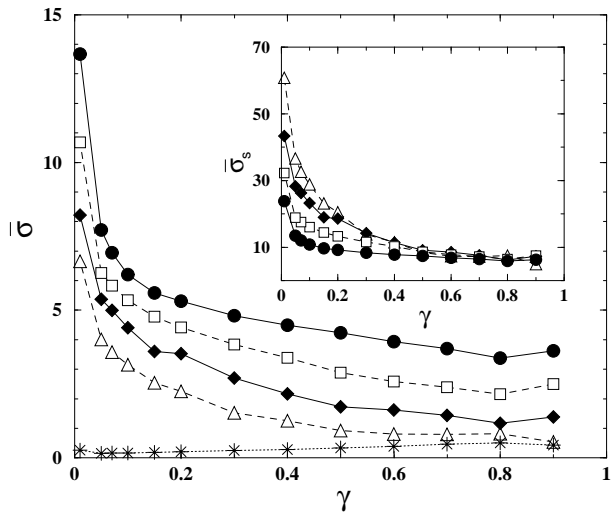}}
\caption{Average dispersion $\overline{\sigma}$ as a function
of $\gamma$, for $K=7$, $\hbar=0.33$ (circles), $0.11$ (squares),
$0.036$ (diamonds) and $0.012$ (triangles).  
Stars show the same quantity for the integrable case
$K=0.7$, at $\hbar=0.012$. 
Inset: scaled dispersion 
$\overline{\sigma}_s=\overline{\sigma}/\sqrt{\hbar}$ 
versus $\gamma$, with same meaning of symbols as in the main figure.}
\label{figure4}
\end{figure}

Finally, we would like to point out that the transition described
here could be observed by means of Bose-Einstein condensates
in optical lattices. A first experimental implementation of the 
kicked rotator model using a Bose-Einstein condensate has been
recently reported \cite{newzeland}. Dissipative cooling techniques
are possible in these systems. Moreover, images of atomic clouds
can be taken, thus measuring their dispersion. 
Also the measured condensate positions should give a clean
kick function (like in Fig.~2) in the case of collapse
and random scattered points in the case of explosion. 
Such experiments would give important information not only on the interplay 
between chaos and dissipation but also on the stability of 
the condensate \cite{raizen2} under the joint effects of chaotic 
dynamics and dissipation.

This work was supported in part by EU (IST-FET-EDIQIP).



\begin{thebibliography}{99}

\bibitem{weissbook}
U. Weiss, {\it Quantum dissipative systems} (2nd Edition), 
World Scientific, Singapore (1999).

\bibitem{ingold}
T. Dittrich, P. H\"anggi, G.-L. Ingold, B. Kramer, G. Sch\"on, and
W. Zwerger, {\it Quantum transport and dissipation}
(Wiley, Weinheim, 1998).

\bibitem{lindblad}
G. Lindblad, Commun. Math. Phys. {\bf 48}, 119 (1976);
V. Gorini, A. Kossakowski, and E.C.G. Sudarshan,
J. Math. Phys. {\bf 17}, 821 (1976).

\bibitem{graham}
T. Dittrich and R. Graham,
Annals of Physics {\bf 200}, 363 (1990).

\bibitem{schack}
T.A. Brun, I.C. Percival, and R. Schack,
J. Phys. A {\bf 29}, 2077 (1996).

\bibitem{brun}
T.A. Brun, Am. J. Phys. {\bf 70}, 719 (2002).

\bibitem{dalibard}
J. Dalibard, Y. Castin, and K. M\o lmer, Phys. Rev. Lett.
{\bf 68}, 580 (1992).

\bibitem{linear}
J. Halliwell and A. Zoupas, Phys. Rev. D {\bf 52}, 7294 (1995).

\bibitem{schack2}
R. Schack, T.A. Brun, and I.C. Percival,
J. Phys. A {\bf 28}, 5401 (1995).

\bibitem{percival}
I.C. Percival, J. Phys. A {\bf 27}, 1003 (1994).

\bibitem{berman}
G.P. Berman and G.M. Zaslavsky,
Physica A {\bf 91}, 450 (1978).

\bibitem{chirikov81}
B.V. Chirikov, F.M. Izrailev, and D.L. Shepelyansky,
Sov. Scinet. Rev. (Gordon \& Bridge), {\bf 2C}, 209 (1981);
Physica D {\bf 33}, 77 (1988); D.L. Shepelyansky,
Dokl. Akd. Nauk (SSSR) {\bf 256}, 586 (1981)
[Sov. Phys. Dokl. {\bf 26}, 80 (1981).

\bibitem{izrailev}
F.M. Izrailev,
Phys. Rep. {\bf 196}, 299 (1990).

\bibitem{raizen}
F.L. Moore, J.C. Robinson, C.F. Bharucha, B. Sundaram, and M.G. Raizen,
Phys. Rev. Lett. \textbf{75}, 4598 (1995).

\bibitem{zaslavsky}
R.Z. Sagdeev, D.A. Usikov, and G.M. Zaslavsky, 
{\it Nonlinear Physics}, Harwood Acad. Pub., NY (1988).

\bibitem{Husimi}
The Husimi function is obtained from smoothing
the Wigner function on the scale of the Planck constant, see e.g.
S.-J. Chang and K.-J. Shi, Phys. Rev. A \textbf{34}, 7 (1986).

\bibitem{noted}
We note that 
$\langle p \rangle_t = \langle x \rangle_t -\langle x \rangle_{t-1} $ 
can be determined from average positions.

\bibitem{noise}
Note that this difference is hardly visible in 
the Poincar\'e sections of Fig.~\ref{figure1}, drawn on a scale
much larger than the Planck cell.  

\bibitem{habib}
T. Bhattacharya, S. Habib, and K. Jacobs,
Phys. Rev. Lett. {\bf 85}, 4852 (2000).

\bibitem{milburn}
A.J. Scott and G.J. Milburn, 
Phys. Rev. A {\bf 63}, 042101 (2001)

\bibitem{deutsch}
S. Ghose, P. Alsing, I. Deutsch, T. Bhattacharya, and S. Habib,
Phys. Rev. A {\bf 69}, 052116 (2004).

\bibitem{newzeland}
G.J. Duffy, S. Parkins, T. M\"uller, M. Sadgrove, R. Leonhardt, and
A.C. Wilson, Phys. Rev. E {\bf 70}, 056206 (2004).

\bibitem{raizen2}
C. Zhang, J. Liu, M.G. Raizen, and Q. Niu,
Phys. Rev. Lett. {\bf 92}, 054101 (2004).

\end{thebibliography}
\end{document}